\documentstyle[preprint,revtex]{aps}
\begin{document}
\preprint{UM-P-92/114}
\preprint{OZ-92/37}
\preprint{ 1992}
\begin{title}
On the CP-odd Nucleon Potential
\end{title}
\author{V.P. Gudkov, Xiao-Gang He and Bruce H.J. McKellar}
\begin{instit}
Research Center for High Energy Physics\\
School of Physics\\
University of Melbourne \\
Parkville, Vic. 3052 Australia
\end{instit}
\begin{abstract}
The CP-odd nucleon potential for different models of CP violation in the one
meson exchange approximation is studied. It is shown that the main contribution
 is due to the $\pi$-meson exchange which leads to a simple one parameter
CP-odd nucleon potential.
\end{abstract}
\newpage
\section {Introduction}

There has been a lot of activity in looking for CP violation in
nuclear\cite{nucl}
 and
atomic systems\cite{krip} in the last few years due to possible large
enhancement factors
for CP violating effects in these systems. The theoretical calculations are
very complicated. To calculate various CP violating
effects in nuclear and atomic systems we need to go through different levels of
theoretical models.  Firstly one needs to obtain the effective low energy CP
violating Lagrangian at the quark level for particular models of CP violation.
The second step is to calculate the CP violating nucleon-nucleon
interaction using
the low energy Lagragian obtained  from the first step. The third step is to
calculate nuclear CP violating effects using the calculated CP violating
nucleon-nucleon interaction and particular nuclear models. And in the case of
atomic systems, one should use particular atomic models to calculate atomic
parameters and take into account nuclear CP-odd moments. Moreover in atomic
systems, CP-odd effects involving electrons must also be considered.
 Each step of these
calculations is model dependent. Even there are methods to exclude or
minimize the dependence on models\cite{gud}, the calculations are still
complicated.

The purpose of this paper is to consider the calculation of the CP
violating nucleon-nucleon interaction using the one meson
exchange approximation. It is well known that in order to describe the
standard P-odd and CP-even nucleon-nucleon
interaction it is necessary to calculate at least six different meson-nucleon
coupling constants\cite{ddh}.
Due to the model dependent nature (QCD at long distances)
of the calculation for each constant and the
different isospin properties and masses
 of the mesons under consideration, the six parameter P-odd and CP-even nucleon
potential leads to difficulties in the calculations of nuclear effects and
a sensitive dependence on nuclear models.

We will show that for the case of CP violation, the structure of the CP-odd
nucleon potential is much more simpler. With good accuracy it is enough
to take into account only the  $\pi$-meson contribution. In other words, we can
use only one parameter (as a consequence a simple structure of the potential)
to calculate the CP violating effects in nuclei and atoms.

This paper is arranged as following.
In section II, we will discuss the reason for the enhancement of the CP
violating $\pi$-meson nucleon interaction for different gauge models of
CP violation. The structure of CP-odd nucleon potential is discussed in
section III. In section IV we give our conclusions.

\section {CP-odd meson-nucleon coupling constrants}

The consideration of different renormalizable CP violating models leads to the
obvious conclusion that there are several different sourses of CP violation:
\begin{itemize}
\item[a.]  Complex quark mass matrices. In the mass eigenstate basis,
there will be CP
violation in the charged current due to exchange of gauge particles. One of the
best know example is the Kobayashi-Maskawa model\cite{km} (the Standard Model).
 The phase of the
determinant of the quark mass matrices is also related to the CP violating
$\theta$-term in QCD.
\item[b.] Complex mixing angles for gauge bosons. An example is
the Left-Right symmetric model\cite{lr}.
\item[c.] Complex vacuum expectation
values of Higgs bosons, for example the Weinberg model\cite{wein}.
\item[d.] CP-odd pure gluonic interaction, and as the
$\theta$-term in QCD\cite{theta}.
\end{itemize}
In a specific process, some or all of
these CP violating sources will contribute. In this section we will discuss
CP-odd meson-nucleon coupling constants which would arise
from all possible contributions.

CP-odd effects in nuclear (low energy) physics  are model dependent in
general. To estimate these effects, the low energy effective Lagrangian for
different models should be used. These effective Lagrangians will include
CP-odd pure quark,  quark-gluon and pure gluonic operators. We will consider
operators up to dimension six. The pure quark
operators will appear in the form current$\times$ current due to gauge boson
exchange, or pseudo-scalar$\times$ scalar structure due to scalar boson
exchange. The most important feature of these Lagrangians is the presence of
the
right-current$\times$ left-current or the pesudo-scalar$\times$ scalar
structures. These operators have enhanced contributions to the CP-odd
pseudo-scalar meson-nucleon couplings. This is a principal difference compared
to the structure of the P-odd and CP-even effective Lagrangian. To describe
the P-odd and CP-even nucleon-nucleon
 interaction, it is necessary to calculate at least
six meson-nucleon coupling constants\cite{ddh}.
We will show in the following that, due
to the enhanced CP-odd pseudo-scalar meson-nucleon coupling
mentioned above it is possible to describe CP-odd nucleon-nucleon
 interaction using
only one meson-nucleon coupling constant.

Let us consider the low energy effective Lagrangian involving
 only u and d quarks.
Exchanging  gauge bosons at the tree level in the a- and b-type of models
will produce the following
structure of the Lagrangian
\begin{eqnarray}
{\cal L}&\sim& L\times L + L\times R + R\times L + R\times R\nonumber\\
&=& C_{LL}O_{LL} + C_{LR}O_{LR} + C_{RL}O_{RL} + C_{RR}O_{RR}\;,
\end{eqnarray}
where $ O_{LL} = \bar u_L \gamma_\mu d_L
\bar d_L \gamma_\mu u_L$ and other operators
are defined in a similar way.  Note that only $L\times R$ and $R\times L$
have CP violating interaction. In the SM
$C_{LR}$ and $C_{RL}$ are zero at the tree and one-loop levels. However
at higher orders they can be generated.

At the tree level the c-type of models will lead to the CP
violating effective Lagrangian
\begin{equation}
{\cal L} \sim  S\times P = C_{SP}\bar q _1 q_2\bar q_3\gamma_5 q_4 + h.c.\;,
\end{equation}
where $q_i$ can be $ u$ and $d$ quarks depending whether a charged or neutral
scalar is exchanged to produce the effective Lagrangian. Again in the SM
$C_{SP}$ can only be generated at higher orders.

The $L\times R$ term also contains a term proportional to $S\times P$. This can
be seen by making a Fierz transformation on $O_{LR}$. We have
\begin{eqnarray}
 O^F_{LR} = -2[ {1\over 3}\bar u_L u_R\bar d_R d_L +
 {1\over 2} \bar u_L \lambda^a u_R \bar d_R \lambda^a d_L] + h.c.  \;.
\end{eqnarray}

We can now clculate the CP-odd meson-nucleon coupling constants for $\pi$-
and $\rho$- mesons from the $L\times R$ term.  Using the factorization
approximation and the vector meson dominance hypothesis, we have
\begin{eqnarray}
\bar g _{\pi^- NN}&\approx& <\pi^- p|L\times R|n>_{CP} = i{ImC_{LR}\over 2}
<\pi^-|\bar d \gamma_\mu\gamma_5 u|0><p|\bar u \gamma_\mu d|n>\nonumber\\
&=& i{Im C_{LR}\over 2}{m_d^2 -m_u^2 \over m^2_\pi} <\pi^-|
\bar d \gamma_5 u |0><p|\bar u d |n>\;,\nonumber\\
\bar g_{\rho NN} &\approx& <\rho^- p|L\times R|n>_{CP} = {Im C_{LR}\over 2}
{m^2_\rho\over f_\rho}g_A\;.
\end{eqnarray}
where $m_\rho$ and $f_\rho$ are the mass and strong form factor for
$\rho$-meson with $f^2_\rho/4\pi \approx 2$,
$g_A$ is the nucleon axial form factor.

Using the Fierz transformed operator $O^F_{LR}$ and the factorization
approximation, we obtain
\begin{eqnarray}
\bar g _{\pi^0 NN} &\approx& {1\over 3} Im C_{LR}
(<\pi^0|\bar d \gamma_5d|0><N|\bar  u u|N> - <\pi^0|\bar u \gamma_5 u|0>
<N|\bar d d|N>)\;.
\end{eqnarray}
{}From the above equation we clearly see that there is a supression factor
$(m^2_d -m_u^2)/m_\pi^2$ for $\bar g _{\pi^- NN}$ compared with $\bar g_{\pi^0
NN}$.

To compare the contributions from the $\pi$ and $\rho$ meson
exchanges to the CP-odd nucleon potential we remind the readers that
for the standard P-odd and CP-even interaction $L\times L$, we have the same
structure of the corresponding coupling constants
\begin{eqnarray}
g^p_{\pi^- NN} &\approx& {C_{LL}\over 2}{{m^2_d - m^2_u}\over m^2_\pi}
<\pi^-|\bar d \gamma_5 u|0> <p|\bar u d |n>\;.\nonumber\\
g^p_{\rho NN} &\approx& {C_{LL}\over 2} {m^2_\rho \over f_\rho}g_A\;.
\end{eqnarray}
It is well know that for the
P-odd and CP-even nucleon potential the contributions
from the $\pi$ and $\rho$ mesons have the same order of magnitude if the
relative strength of the couplings is given by eq.(6)\cite{ddh}.
It is expected that
the same thing should happen for the CP-odd nucleon potential. From eq.(4)
we expect that the contributions from the $\rho$ and $\pi^-$ meson
exchanges to the CP-odd potential will have the same order of magnitude.
Therefore we conclude that the dominant contribution to
the CP-odd nucleon potential is from the $\pi^0$ meson exchange.

We have similar results for the c-type of models. In this type of models, the
$\rho$-meson nucleon coupling will be much smaller than $\pi$-meson nucleon
couplings for the same reason given above. However unlike the situation in
the a- and b- type of models where the $\pi^0$
meson-nucleon coupling is much larger than the $\pi^-$ meson-nucleon
coupling, the
charged and neutral pion-nucleon coupling can be the same order of magnitude.
Therefore $\pi^\pm$ and $\pi^0$ exchange can all make significant contributions
to CP-odd nucleon potential.
The reason for this enhancement on $\bar g_{\pi NN}$ is due to the large
contribution of the pseudo-scalar and scalar quark densities in the local
approximation. A similar enhancement factor for the strange quark current has
been found in penguin induced K-meson decays\cite{zvsh}.
We conclude that the dominant CP violating
nucleon-nucleon interaction in the one meson exchange approximation is from
the $\pi$-meson exchange. The situation here is quite different from the
P-odd and CP-even nucleon potential where the $\pi$, $\rho$ and $\omega$ all
contribute significantly\cite{ddh}.

CP-odd pure gluonic operators ($J^{PC} = 0^{-+}$) can be generated
in many models\cite{mor}, in particular the c- and d-type of models.
It is interesting to
note that because of the pseudo-scalar nature of the operators,
the pseudo-scalar meson-nucleon
coupling constants are much bigger than the vector meson-nucleon coupling
constants, just as they are for the pure quark operators. To estimate the
ratio of the coupling constants for pseudo-scalar
and vector mesons we will use two different methods: 1) using reduction
formulas, and 2) using naive dimensional analysis. For the first mehtod,
in the low
energy limit, we obtain for the meson nucleon matrix elements
\begin{eqnarray}
<\rho N|\hat O |N> \approx i m_\rho^2\int dx e^{-iqx} <N|T\{\hat O(0),
\rho(x)\}|N>\;,\nonumber\\
<\pi N|\hat O|N> \approx {i\over f_\pi}\int dx e^{-iqx} <N|T\{\hat O(0),
\partial_\mu A_\mu\}|N>\;.
\end{eqnarray}
where $\hat O(x)$ is a CP-odd pseudo-scalar gluonic operator,  $f_\pi$ is
the pion decay constant.  In the last formula
we have used the PCAC hypothesis to express the pseudo-scalar field in
terms of a
axial vector current. Using the vector meson dominance hypothesis,
one has $\rho_\mu =
(m^2_\rho/f_\rho)V_\mu$ with $V_\mu$ being a vector current.
We now use the factorization method to estimate the nucleon matrix elements
\begin{equation}
<N|T\{\hat O, \hat A\}|N> \approx <0|\hat O|0><N|\hat A|N>\;.
\end{equation}
{}From this we obtain
\begin{equation}
{<\pi N|\hat O|N>\over <\rho_\mu N|\hat O|N>} \approx {M_N\over f_\pi g_\rho}
\gg 1\;,
\end{equation}
where $g_\rho = f^2_\rho /4\pi $.

The naive dimensional analysis gives
\begin{equation}
{<\pi N|\hat O|N> \over <\rho_\mu N|\hat O|N>}
\approx {g_{\pi NN}\over g_\rho} \;.
\end{equation}
where $g_{\pi NN}$ is the strong $\pi NN$ coupling constant. Using the
Goldberger-Treiman relation, we can see that this result conrresponds to
the estimate of eq.(9). Therefore, we conclude that for gluonic CP-odd
operators
the coupling constant of the pseudo-scalar meson to nucleon is larger by about
one order of magnitude than vector meson-nucleon coupling constant.

Let us now discuss operators with quarks and gluons. The lowest order
CP-odd quark-gluon operator is the colour-electric dipole moment
\begin{equation}
\hat O = \bar q \sigma _{\mu\nu}\gamma_5 {\lambda^a\over 2} q G^{a\mu\nu}\;.
\end{equation}
We can apply the same argument as for the pure gluonic operators to conclude
that one will obtain a large $\bar g_{\pi NN}/\bar g_{\rho NN}$ ratio from
this operator.  Alternatively, we can use the extended relation
\begin{equation}
<g_s\bar q \sigma_{\mu\nu}\gamma_5 {\lambda^a\over 2}q G^a_{\mu\nu}>
\approx m^2_{st}<i\bar q \gamma_5 q>\;.
\end{equation}
in analogy to the well know relation $<ig_s\bar q \sigma_{\mu\nu}
{\lambda^a\over 2} q G^a_{\mu\nu}> = m^2_0<\bar q q>$\cite{yo}.
Here $m_0$ and $m_{st}$
are characteristic masses about 0.9 GeV. Together with the use of the PCAC
hypothesis we obtain similar results to eq.(10). The analysis for the
$\rho$ meson contribution can also be applied to the $\omega$ meson.

Now we can give the main result of this section. For all types of models
of CP violation, in the one meson exchange approximation the contributions
to the CP-violating nucleon-nucleon interaction from
pseudo-scalar mesons are larger than the constributions from vector meson
by about one order of magnitude. Therefore, to calculate CP-odd effects in
nuclei with a reasonable accuracy, we need only consider
pseudo-scalar meson exchange.

\section{CP-odd nucleon potential}

To describe the CP-odd nucleon potential due to one meson exchange we restrict
oursleves by mesons with masses less than $\rho$-meson mass, since
contributions
 from heavy mesons are suppressed in low energy limit ($E\leq E_{fermi}$) due
to
the repulsive force (due to the nucleon core). We will need to consider
contributions from the  $\pi$, $K$, $\eta$, $\rho$ and $\omega$ mesons. We have
 argued in the previous section that the $\rho$ and $\omega$ contributions are
much smaller than that of the pseudo-scalar mesons. We can safely neglect their
 contributions and only consider the pseudo-scalar contributions.
In the  b-, c- and d-type of  models ,
the CP-odd meson-nucleon coupling will be generated at the first order in weak
interaction. If only nucleon-nucleon interactions between neutron and proton
are concerned, the $K$ meson exchange will not contribute to the lowest order.
For these types of models we only need to consider the $\pi$ and $\eta$
contributions. It is obvious that the nucleon potentials for both mesons have
the same spatial behaviours. We have\cite{hax}
\begin{eqnarray}
V_{CP}^{\pi} = -{m_\pi^2 \over 8\pi m_N} g_{\pi NN} \bar g_{\pi NN}
(\vec \tau_1 \cdot \vec \tau_2)(\vec {\bf \sigma}_1 - \vec {\bf \sigma}_2)\cdot
\vec {\bf r}
{e^{-m_\pi r}\over m_\pi r^2}[ 1 + {1\over m_\pi r}]\;,\nonumber\\
V_{CP}^{\eta} = -{m_\eta^2\over 2\pi m_N}g_{\eta NN}\bar g_{\eta NN}
(\vec {\bf \sigma}_1 - \vec {\bf \sigma}_2)\cdot {\bf \vec r}
{e^{-m_\eta r}\over
m_\eta r^2}
[1 + {1\over m_\eta r}]\;,
\end{eqnarray}
where $g$ and $\bar g$ are CP-even and CP-odd coupling constants,
${\bf \vec \sigma}$
and ${\bf \vec \tau}$ are the spin and isospin of the nucleon, respectively.

Using the expressions for the CP-odd nucleon potentials, one can estimate the
relative contributions of them to CP-odd nucleon interactions. It should be
mentioned that such estimation is natural for systems with large number
of nucleons, but may not be valid for a few nucleon systems where some CP-odd
effects are forbiden by isospin selection rules. The ratio of $\bar g_{\pi NN}/
\bar g_{\eta NN}$ is usually less than $m^2_\eta /m^2_\pi$, therefore one
obtains for short distance interaction ($r\approx 1/m_\rho$)
\begin{equation}
{V^\pi_{CP}\over V^\eta_{CP}} \sim 1\;,
\end{equation}
and for large distance interaction ($r \approx 1/m_\pi$)
\begin{equation}
{V^\pi_{CP}\over V^\eta_{CP}} \sim 10\;.
\end{equation}

Since almost all nuclear effects have the main contribution from the long
distance region ($r\sim 1/m_\pi$), we conclude that for the calculation of
CP-odd nuclear effects with a reasonable accuracy ($10\%$)
it is enough to know only one
CP-violating constant: the $\pi$-nucleon coupling constant.

In the SM CP-odd meson-nucleon couplings can be generated only
at the second or higher orders
in weak interaction. In this case besides the contributions to the
CP-odd nucleon potential from the $\pi$-mesons, the K-mesons also have
significant contributions. This has been discussed in detail in Ref.\cite{hm}.

The coupling constants of $\pi$ and $\eta$ to nucleon for other types of
 models of
CP violation have been calculated by several authors\cite{gud,hm,coup}.
Using these results,
it is possible to calculate the electric dipole moments of nuclei\cite{krip}
 and CP-odd
effects in nucleon scattering\cite{nucl}.
The simple structure of the CP-odd nucleon
potential gives the unique opportunity to calculate CP-violating effects in
complicated nuclei.

\section{Conclusions}
The nucleon CP-odd potential has main contribution from one $\pi$-meson
exchange for various models of CP violation. Therefore for the estimation of
different CP-violating effects in nuclear physics it is necessary to calculate
only the $\pi$-meson nucleon coupling constant. This fact leads to the simple
parametrization of all CP-odd effects using only one parameter, and provides
an oppotrtunity to test different models of CP violation.

Using one parameter  CP-odd $\pi$-nucleon coupling, it is possible to obtain
direct relation between CP-odd nuclear effects and the value of the neutron
electric dipole moment if the one meson loop gives the main contribution\cite
{theta,hm,hm2}

\acknowledgments
This work is supported in part by the Australian Research Council.

\end{document}